\documentclass[amsart]{article}
\usepackage{authblk}
\usepackage{graphics}
\usepackage{caption}
\usepackage{mathtools}
\usepackage[flushleft]{threeparttable}
\usepackage{amsmath}
\usepackage{lineno}
\usepackage{mathrsfs}
\usepackage{graphicx}
\usepackage{wrapfig}
\usepackage{epstopdf}
\usepackage{enumitem}
\usepackage{lipsum}
\usepackage{placeins}
\usepackage[margin=1in]{geometry}
\usepackage{abstract}
 
\title{Origin of Negative Temperatures in Systems Interacting with External Fields}
\author[1]{Salvatore Calabrese}
\author[1,2,*]{Amilcare Porporato}

\affil[1]{Department of Civil and Environmental Engineering, Princeton University, Princeton, NJ, USA.}
\affil[2]{Princeton Environmental Institute, Princeton University, Princeton, NJ, USA.}
\affil[*]{Corresponding author: Amilcare Porporato, aporpora@princeton.edu}
\begin{document}
	\maketitle
	
\section*{Abstract}
The controversial existence of negative temperatures has stirred interesting debates that have reached the foundations of thermodynamics, including questions on the second law, the Carnot efficiency and the statistical definition of entropy. Here we show that for systems interacting with an external field, negative temperatures arise from an energy mis-attribution in which the interaction energy with the field is treated as a form of internal energy. We discuss how negative temperatures are avoided when using a proper thermodynamic formalism, which accounts for the intensive and extensive variables associated to the external field. We use the paramagnetic system and a perfect gas in a gravitational field to illustrate these ideas. Considerations about the isothermal and adiabatic work done by the field or the system also shed light on the inconsistency of super-Carnot efficiencies.

\section{Introduction}	

Negative absolute temperatures have been introduced in systems where the density of states is a locally decreasing function of energy {\cite{ramsey1956thermodynamics,lavenda1999donegative,braun2013negative,hilbert2014thermodynamic,frenkel2015gibbs}}. Examples include spin systems {\cite{purcell1951nuclear,hakonen1994negative,pathria2011statistical}}, lasers {\cite{yariv1989quantum,zhukov1997negative}}, and systems with long-range interactions {\cite{campa2009statistical,touchette2015equivalence}}. In an early review, Dyson {\cite{dyson1954heat}} describes negative temperatures as 'just a philosophical curiosity'; other authoritative references interpret them as the result of convenient conventions {\cite{kestin1979course}} and specify that negative temperatures are actually higher than positive ones {\cite{landau1969statistical,pathria2011statistical}}, in the sense that heating from absolute zero to +$\infty$ K, which is identical with -$\infty$ K, continues from there to -0 K {\cite{kestin1979course}}. Besides mostly interpretative issues, one particularly troubling point with negative temperatures is the emergence of super-Carnot efficiencies when performing thermodynamic cycles between positive and negative temperatures {\cite{frenkel2015gibbs}}.

Lavenda {\cite{lavenda1999donegative}} was perhaps the first author to raise objections against negative temperatures. He noticed that the crux of the matter is whether the magnetic energy is a form of heat and to avoid negative temperatures he proposed a modification of the second law. Often the discussion about negative temperatures centers around spin systems, in which the sudden reversal of the external field induces a population inversion that, while it persists, is characterized by an extensive variable that is opposite in sign to the field \cite{purcell1951nuclear,ramsey1956thermodynamics}. If one interprets them as equilibrium states, the difference in orientation between the field and the extensive variable is resolved with a negative temperature. As noted in \cite{struchtrup2018work}, these states are unstable leading to ambiguities in the definition of thermodynamic temperature.

In parallel, a series of papers focusing on isolated systems \cite{sokolov2014thermodynamics,hilbert2014thermodynamic,campisi2015construction} also disputes their existence, basing their criticism on shortcomings of the Boltzmann entropy, and advocates for other forms of microcanonical entropy founded on Gibbs volume entropy. The latter, being proportional to the total number of microstates per energy lower than a given energy, is in fact always a monotonically increasing function of the energy and thus cannot give rise to negative temperatures. However, the fact that Gibbs entropy clashes with some basic principles of thermodynamics \cite{frenkel2015gibbs,swendsen2015gibbs} has prompted a defense of negative temperatures by other authors {\cite{frenkel2015gibbs,poulter2016defense}}. The ensuing debate has spurred a welcome discussion of the definition of entropy for small systems {\cite{hilbert2014thermodynamic,campisi2015construction,swendsen2016negative}}, however, being focused on isolated systems does not address the role of interactions with the environment \cite{kirkwood1935statistical,jarzynski2017stochastic}. Furthermore, in these works the existence of decreasing branches in the density of states of truly isolated system is assumed a priori and the discussion centered on theoretical grounds. To this regard, it is useful to recall Callen's warning (\cite{callen2006thermodynamics}, Sec. 15-3) that the decreasing branch in the density of states of the two states model is outside the range of energies for which the model should be used.

Here we bypass the issue of the possible definitions of entropy (and the related matters on classical and quantum statistics and the equivalence of the ensembles), which is mainly relevant to small systems, to focus on systems in the thermodynamic limit. We show that when thermodynamic systems interact with an external field, such as magnetic or gravitational, negative temperatures emerge from an energy-entropy mis-attribution. On the contrary, the adoption of a correct thermodynamic formalism, which includes the work done by such fields (\cite{callen2006thermodynamics}, p. 81-83), leads to intensive quantities that can be negative but do not correspond to the true thermodynamic temperatures measurable with a thermometer. The ordering effect brought about by the work done on the system by an external field \cite{landsberg1994entropy} is linked to a reduction in the number of particular microstates and a corresponding lowering of the entropy. As a result, the field force appears as an entropic force whose isothermal work done to order the states of the system ends up being expelled into the environment as heat, in a condition which is reminiscent of the famous Landauer erasure principle {\cite{landauer1961irreversibility,parrondo2015thermodynamics}}.

After a general theoretical discussion of the thermodynamics and statistical mechanics in the presence of an external field, we illustrate the details of our argument by means of two specific models: a paramagnetic system, which is typically used to explain negative temperatures, and a perfect gas within horizontal plates in a gravitational field. For such models, the calculations can be carried out in detail with simple methods and without distractions from the physical interpretation.

\section{\label{sec:theory} Thermodynamics of systems in an external field}

The Gibbs relation for a simple system \cite{simplesystems} (no external fields) with internal energy  $U$ and entropy $S$ and constant mass and volume can be written as
\begin{equation}
	\label{eq:gibbssimp}
	dS=\frac{1}{T}dU,
\end{equation}
where $T=\frac{d U}{d S}$ is the thermodynamic temperature and $dQ=TdS=dU$ is the heat exchanged reversibly with the environment. The latter equality reflects the fact that heat is a disordered form of energy and temperature is the intensive quantity associated to it {\cite{dyson1954heat}}. Entropy is a first-order homogeneous function of its argument and the corresponding Euler relation has the form (\cite{callen2006thermodynamics}, p. 60)
\begin{equation}
	S=\frac{1}{T}U+S_0.
\end{equation}
Furthermore, $S$ is a convex, monotonically increasing function of $E$; these properties ensure the stability of the thermodynamic equilibrium and the existence of strictly positive temperatures {\cite{callen2006thermodynamics}}. The entropy $S_0$ accounts for the microstates not directly related to thermal agitation, including those that arise from the spin orientations in a system of particles consisting of magnetic dipoles.

When the system interacts with an external field, $\phi$, the Gibbs relation needs to be extended, as done for the gravitational field by Gibbs and Boltzmann \cite{gibbs1928collected,huang1964statistical,keszei2013chemical}, and shown also by Callen for magnetic systems (\cite{callen2006thermodynamics}, p. 479-483). As a result, in the entropy representation we have \cite{callen2006thermodynamics}
\begin{equation}
	\label{eq:gibbsext}
	dS=\frac{1}{T}dU+\frac{\phi}{T} dX,
\end{equation}
where $X$ is the extensive variable associated to $\phi$, such that $\frac{\phi}{T}=\frac{\partial S}{\partial X}$. In this expression, $T$ is still the thermodynamic temperature, which remains strictly positive, whereas $\frac{\phi}{T}$ can be either positive or negative depending on the sign of the field. Equation (\ref{eq:gibbsext}) expresses the fact that both changes in $U$ and in $X$ affect the entropy of the system, namely $S=S(U,X)$. The presence of the external field brings about an ordering of the system \cite{landsberg1994entropy} and a corresponding reduction of $S_0$. In a sense, this effect is analogous to the one of raising a thermodynamic wall in a composite system {\cite{callen2006thermodynamics}}. Moreover, when $X$ has units of length, $\phi$ has units of force and can be interpreted as a thermodynamic force associated to the entropy reduction by the field.

Negative temperatures arise when the term $\phi X$ is considered as part of the internal energy. This implicitly happens when the system is improperly treated as a simple system (in the sense of thermodynamics \cite{simplesystems}) using an overall energy $U'=U+\phi X$ in (\ref{eq:gibbssimp}), i.e.,
\begin{equation}
	\label{eq:gibbssimp2}
	dS=\frac{1}{T}dU+\frac{\phi}{T} dX=\frac{1}{T'}dU'.
\end{equation}
Since $T'$ is not the thermodynamic temperature, it should not be expected to be strictly positive. Indeed, $T'$ is equal to the thermodynamic temperature $T$ only for iso-$X$ processes, i.e., $\frac{1}{T'}dU'=\frac{1}{T}dU $, and it becomes negative for processes at constant $U$, $\frac{1}{T'}=\frac{\phi}{T}$, when the external field is negative (e.g., upon inversion of a magnetic or gravitational field). This mis-attribution is analogous to confusing the enthalpy, $H=U+pV$, with the internal energy, $U$, by including the Landau potential, $pV$, as part of the internal energy.

In a spin system, for example, the extensive parameter $X$ corresponds to the magnetization and its conjugate intensive parameter is the magnetic field. Their product gives a potential and is the energy arising from the interaction with the magnetic field. As we also show later on, for a correct thermodynamic formalism this energy should be explicitly accounted for in the Gibbs relation \cite{callen2006thermodynamics} (p. 479-483).

From the point of view of statistical mechanics, the system intensive variables can be computed from the Boltzmann postulate, $S=k \ln \Omega$, where $k$ is the Boltzmann constant and $\Omega(U,X)$ is the number of microstates,
\begin{equation}
	\frac{1}{k T}= \frac{k}{\Omega}\frac{\partial \Omega}{\partial U} \quad {\rm and} \quad \frac{\phi}{k T}= \frac{k}{\Omega}\frac{\partial \Omega}{\partial X}.
\end{equation}
In this context, it is crucial to recognize that the number of microstates depends on $X$. Otherwise a fictitious temperature $T'$ appears again when treating $\Omega$ as a function of $U'=U+\phi X$ and computing the entropy as $S(U')=k \ln \Omega_{U'}$, that is
\begin{equation}
	\frac{1}{k T'}=\frac{k}{\Omega_{U'}}\frac{\partial \Omega_{U'}}{\partial U'}.
\end{equation}
This is also the case when constructing the thermodynamic ensembles using a Hamiltonian formulation, since for simple systems it is typical to equal the Hamiltonian $\mathcal H$ to the system internal energy, while here it has the additional contribution due to the field, i.e., $\mathcal H=U'=U+\phi X$.

In what follows, we will focus on the simpler case where the equations of state are functions only of their respective variables, i.e., $\partial S/\partial U=1/(kT)=\beta(U)$ and $\partial S/\partial X=\phi/(kT)=\gamma(X)$. This is the case for the ideal gas, where due to the absence of intermolecular forces only the kinetic energy contributes to the internal energy, while the potential energy associated to the infrequent elastic collisions is negligible. Thus, one can alter the extensive variables, $X$ (e.g., the volume), without affecting the internal energy, $U$. For these systems, the density of states can be factorized as $\Omega=\Omega_{U}\Omega_{X}$ \cite{pathria2011statistical}, so that
\begin{equation}
	\label{eq:boltzmannpost3}
	S=k\ln \Omega_{U}+k\ln \Omega_{X}=k\ln \Omega_{U}+S_0+k\ln \frac{\Omega_{X}}{\Omega_0},
\end{equation}
where $S_0=k \ln \Omega_0$; the latter step emphasizes the lowering of entropy due to the reduction of the number of configurations compared to the value of $S_0$ when the external field $\phi$ is zero.

\section{Paramagnetic system}
Magnetic systems provide an emblematic example for the discussion of negative temperatures {\cite{pathria2011statistical}}. We consider an isolated system of $N$ particles with spin-$1/2$ and intrinsic magnetic moment $\mu$. $N^+$ particles have positive spin while the remaining $N^-=N-N^+$ particles have negative spin so that the total magnetization is $M=\mu (N^{+}-N^{-})$. It is assumed that there are $\Omega_M$ equiprobable ways to arrange the spins among the particles consistent with the magnetization $M$ and linked to an entropy $S_M$.
The total entropy of the system,
\begin{equation}
	\label{eq:totentromagn}
	S=S_U+S_M,
\end{equation}
also includes the entropy linked to thermal configurations, $S_U=k \ln \Omega_U$. In what follows, we consider constant internal energy, $U$, and focus on the variation of the entropy with respect to the magnetization of the system, $M$.

\begin{figure}[t]
	\includegraphics[width= 13cm]{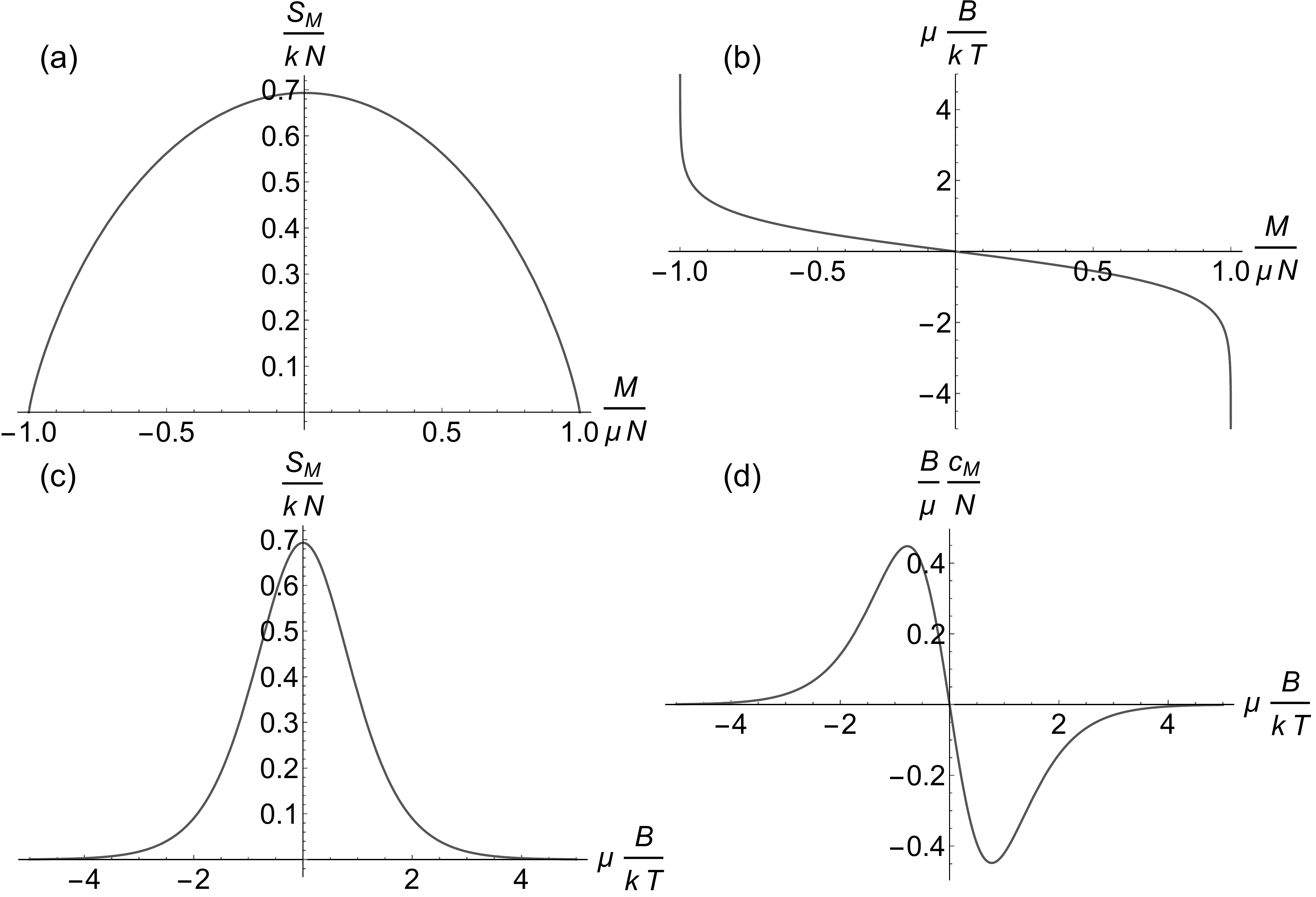}
	\centering
	\caption{\footnotesize (a) Entropy and (b) magnetic field as a function of the normalized magnetic moment $M/\mu N$. (c) Entropy and (c) magnetic susceptibility as a function of the magnetic field $\mu \frac{B}{T}$. The slope of the entropy curve with respect to the magnetization, shown in (b), is not the thermodynamic temperature but the magnetic field divided by the temperature ($B/T$), see equation (\ref{eq:magnfield}).}
	\label{Fig:figure1}
\end{figure}

The number of microstates $\Omega_{M}$ due to the possible spin configurations is {\cite{davidson1962statmechan}}
\begin{equation}
	\label{eq:omeisom}
	\Omega_{M}=\frac{N!}{N^{+}!N^{-}!}.
\end{equation}
From the Boltzmann postulate and invoking the Stirling approximation for large systems, one obtains
\begin{equation}
	S_M=-kN^-\ln \frac{N^-}{N}-kN^+\ln \frac{N^+}{N}
\end{equation}
and, using the definition of $M$ and the fact that $N^-=N-N^+$,
\begin{equation}
	\label{eq:omeisom2}
	S_M=-N k\frac{1}{2}\left(1-\frac{M}{N \mu}\right)\ln \left[\frac{1}{2}\left(1-\frac{M}{N \mu}\right)\right]- N k \frac{1}{2}\left(\frac{M}{N \mu}+1\right) \ln \left[\frac{1}{2}\left(\frac{M}{N \mu}+1\right)\right].
\end{equation}
As seen in Figure (\ref{Fig:figure1}), this entropy is a convex function of the magnetization and is symmetrical about $M/\mu=0$. The macrostate of zero magnetization, with half spins positive and half negative, is the one with the highest number of microstates and thus of maximum disorder. Increasing magnetization increases the order of the spins, until a minimum entropy is reached when all the spins have same orientation.

\begin{figure}
	\includegraphics[width=13 cm]{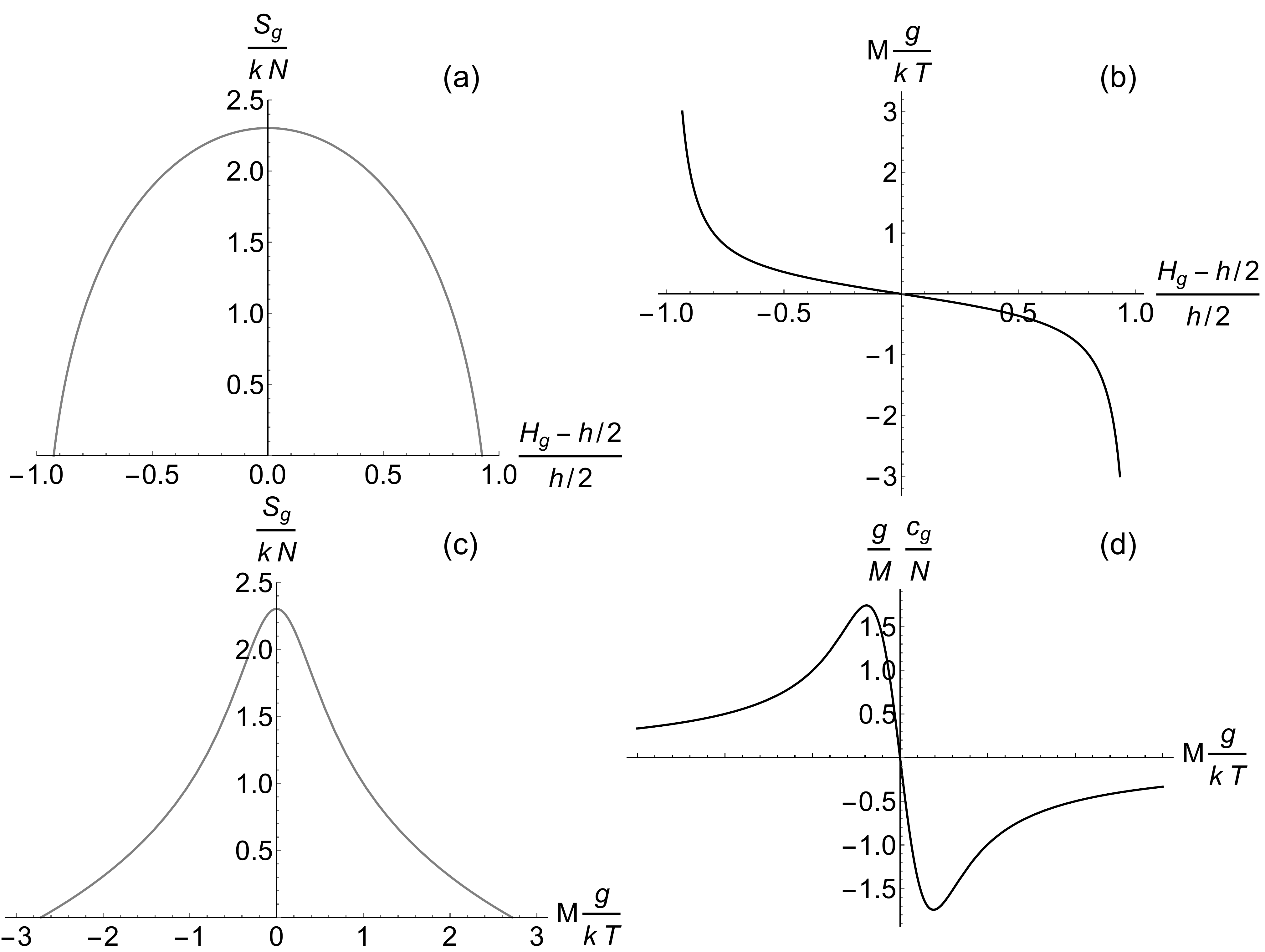}
	\centering
	\caption{\footnotesize (a) Entropy and (b) intensity of the gravitational field as a function of normalized displacement of the center of gravity. (c) Entropy and (d) heat capacity as a function of $M\frac{g}{k T}$.}
	\label{Fig:figure2}
\end{figure}

Having identified the entropy of the system as a function of the macroscopic variables, it is now possible to compute the intensive variables by differentiating (\ref{eq:totentromagn}). The result coincides with equation (B.19) of Callen \cite{callen2006thermodynamics} and reads
\begin{equation}
	\label{eq:gibbsmagn}
	dS=\frac{1}{T}dU+\frac{B}{T} dM,
\end{equation}
where
\begin{equation}
	\label{eq:temp}
	\frac{1}{k T}=\frac{\partial S}{\partial U}=\frac{1}{\Omega_{U}}\frac{\partial\Omega_{U}}{\partial U}
\end{equation}
and
\begin{equation}
	\label{eq:magnfield}
	\frac{B}{k T}=\frac{\partial S}{\partial M}=\frac{1}{\Omega_{M}}\frac{\partial \Omega_{M}}{\partial M}
\end{equation}\\
with $B$ the magnetic field. Using these relationships, it is easy to obtain the entropy as $S_M=S_M(B/T)$, which is plotted in Figure \ref{Fig:figure2}(c)), along with the magnetic susceptibility (Figure \ref{Fig:figure2}(d)),
\begin{equation}
	\label{eq:magnsusc}
	c_M=\frac{\partial M}{\partial B}=\frac{T}{B}\frac{\partial S_M}{\partial B}.
\end{equation}
\\
Equation (\ref{eq:temp}) makes it clear that the thermodynamic temperature $T$ is determined exclusively by the internal energy $U$ and its configurations $\Omega_{U}$, and not by the magnetic energy $B M$. In contrast, the magnetic field $B$ depends only on $M$ and the number of spin configurations $\Omega_{M}$. When most spins are up, $M>0$ and $B<0$, while when most spins are down $M<0$ and $B>0$ (Figure \ref{Fig:figure1}(b)). In the energy representation, $U=T S-B M$ and, since $S>0$ and $B M<0$, then $U>0$.
The magnetic susceptibility is highest at the lowest magnetization and then decreases with respect to $M/\mu$ (see the inverse of the slope of Figure \ref{Fig:figure1}(b)). Equivalently, $c_M$ goes to zero at saturation (e.g., $M=\mu N^+$), whereas it tends to a constant (Curie's law) as $B/T\to 0$, where it becomes a linear function of the inverse temperature with a slope equal to the Curie constant \cite{kittel1996introduction}. If one plots the magnetic susceptibility scaled as $\frac{B}{\mu N} c_M$ with respect to $\mu \frac{B}{T}$, a so-called Schottky anomaly appears because of the particular normalization {\cite{pathria2011statistical}}.

The two state paramagnetic model gives rise to negative temperatures {\cite{ramsey1956thermodynamics,masthay2005positive,pathria2011statistical,frenkel2015gibbs,malgieri2018study}} when the internal energy is identified with the magnetization, for example by giving an energy $\epsilon$ to the positive spins while attributing zero energy to the negative ones and neglecting the thermal contribution. Accordingly, the system 'internal energy' is defined as $U'=\epsilon N^+$ and the entropy is considered only as function of $U'$,
\begin{equation}
	S=kN \ln N-k \left(N-\frac{U'}{\epsilon}\right)\ln \left(N-\frac{U'}{\epsilon}\right)-k \frac{U'}{\epsilon}\ln \frac{U'}{\epsilon}.
\end{equation}
Applying equation (\ref{eq:gibbssimp2}) one then obtains
\begin{equation}
	\frac{1}{k T'}=\frac{d S}{d U'},
\end{equation}
which is positive when $U'<\epsilon \frac{N^+}{2}$, negative when $U'>\epsilon \frac{N^+}{2}$, and has an asymptote at $U'=\epsilon \frac{N^+}{2}$. However, given that one has assumed $dU=0$, $1/T'$ turns out to be nothing but $B/T$, whose sign follows the orientation of the magnetization $M$.

\section{Ideal gas in a gravitational field}

Our second application consists of an ideal gas in a uniform gravitational field $g$ between two horizontal plates at a distance $h$. In the absence of gravity, the molecules of gas uniformly fill the space and both the particle density $\rho$ and pressure $p$ are constant throughout the height $h$. Instead, with a gravitational field of intensity $g$, the molecules are pushed against the plate in the direction of the field. While so far the problem has mainly been approached by means of statistical ensembles {\cite{landsberg1994entropy,roman1997microcanonical}}, here the system is assumed to be made of a very large number of molecules ($\approx 10^{23}$) such that the continuum hypothesis holds and the various thermodynamic quantities can be easily computed by assuming local thermodynamic equilibrium. In isothermal conditions at temperature $T$, this results in the well-known exponential distributions for density and pressure already obtained by Gibbs and Boltzmann \cite{gibbs1928collected,berberan1997barometric},
\begin{equation}
	\label{eq:denspressdistr}
	\rho(z)=\frac{p(z)}{k T}=\frac{p_0}{k T} e^{-\frac{M g z}{k T}},
\end{equation}
where $z$ is the vertical coordinate, $M$ is the molecular mass, and $p_0$ is found by imposing $\int\limits_{0}^{h}\rho(z)dz=N$, $N$ being the total number of molecules. Thus the mechanical equilibrium between pressure and gravitational forces creates an inhomogeneity in density and pressure (and therefore also in the chemical potential), while being in thermal equilibrium at constant $T$. The uniformity of the vertical temperature profile, ensured by the fact that the gas and the plates are assumed dia-thermal, was clarified by Maxwell and Boltzmann in a interesting debate against an objection by Locksmith {\cite{trupp1999energy}}.

With these premises, it is easy to see that the thermal entropy, $s_U=c_v \ln T$, does not depend on $z$ or $g$, while the field-related entropy per unit volume,
\begin{equation}
	\rho s_g(z)=-k\rho(z)\ln\frac{\rho(z)}{N},
\end{equation}
depends both on $z$ and $g$. In turn, the total entropy for the layer of gas is found by integration with respect to $z$
\begin{eqnarray}
	\label{eq:entridealgas}
	S=S_U+S_g= \frac{k N}{R} c_v\ln T-\int\limits_{0}^{h}k\rho(z)\ln\frac{\rho(z)}{N}dz=\\
	\frac{k N}{R} c_v\ln T+k N \left(1-\ln \varphi -h \varphi e^\gamma \right),
\end{eqnarray}
in which $R$ is the gas constant, $\gamma=g h M/kT$, and $\varphi= \gamma/(h (e^\gamma-1))$. Thus the global entropy (\ref{eq:entridealgas}) depends on $g$ and, as for the spin system, is maximized in the absence of the external field, $g\to 0$, and decreases as the gravitational field is increased, regardless of its direction (see Figure \ref{Fig:figure1}(c)). As already noticed by Landsberg et al. \cite{landsberg1994entropy}, beyond a certain $|g|$ the entropy becomes negative. This is simply due to the fact that locally the density is too large for the gas to be still treated as ideal {\cite{streater1994convection}}.

To formulate an extended Gibbs equation for the overall system analogous to (\ref{eq:gibbsmagn}) for the spin system, it is necessary to find the extensive variable $X$ associated to $g/T$. From the fact that $\partial S/\partial X=g/T$, we have $\partial S/\partial g=g/T (\partial X/\partial g)$ and by integration
\begin{equation}
	X=\int\frac{T}{g}\frac{\partial S}{\partial g}dg.
\end{equation}
Substituting (\ref{eq:entridealgas}) in the previous expression gives $X=N M H_g=m H_g$, where $m$ is the total mass and $H_g=(\int_0^h z \rho dz)/N $ is the center of gravity of the gas. As a result, the Gibbs equation can be written as \cite{keszei2013chemical}
\begin{equation}
	\label{eq:gibbsgrav1}
	dS=\frac{1}{T}dU+\frac{g}{T} m dH_g.
\end{equation}
By analogy with (\ref{eq:magnsusc}), the gravitational susceptibility $c_g$ is defined as $c_g=NM dH_g/dg$.

These results, summarized in Figure (\ref{Fig:figure2}), bear a striking similarity to those of the paramagnetic model. They show that the gravitational field tends to organize the system by displacing the center of gravity in the direction of the field. In the limit $|g| \to \infty$  the system is squeezed towards the lower or upper plate depending on the sign of $g$ and $S_g \to -\infty$, while for $|g| \to 0$ the density profile is uniform and the entropy $S_g$ reaches its maximum. Analogously to the magnetic system, also in this case, when the external field (i.e., the gravitational energy) is considered as part of the internal energy in (\ref{eq:gibbssimp2}), one obtains a negative temperature as soon as the gravitational potential is inverted. In this case, however, the macroscopic and classical nature of the system leaves little doubt that such negative temperatures only have a formal meaning and do not correspond to real temperatures.

\section{Conclusions}

Negative temperatures originate from a mis-attribution of energy and disappear when properly accounting for the ordering effects of and the related work done by the external field. Negative slopes $\frac{\partial S}{\partial M}=\frac{B}{T}$ in a paramagnetic systems are linked to a negative direction of the magnetic field, rather than to a real inversion of the absolute temperature scale. The case of an ideal gas confined vertically by two parallel plates reproduces the same thermodynamic idiosyncrasies of the magnetic system, when the thermal configurations are not carefully distinguished from those affected by the external field.

Moving along the entropy curve in Figure \ref{Fig:figure1}(a) and \ref{Fig:figure2}(a) corresponds to a reversible isothermal process in the extended Gibbs equations (\ref{eq:gibbsext}), in which the work done by the external field $\Delta w=\int \phi dX $ is released by the system as heat $T \Delta S$ into the environment. For the two applications considered here, the independence between configuration types also implies $\Delta U=0$ for an isothermal process, analogously to an isothermal compression/expansion of a gas. Thus, in isothermal transformations when the external field is turned on and work is done to order the system, heat is released to maintain constant temperature, while when the external field is turned off, thermal agitation acts to restore the disorder, i.e., increase the entropy, absorbing heat from the environment. In adiabatic processes, the work done by the external field is stored as internal energy and temperature increases, corresponding to a transfer of configurations from magnetic or gravitational to thermal. Switching off the external field, the system transfers back thermal configurations to magnetic or gravitational modes, reducing the temperature. In this way, super-Carnot efficiencies that appear when introducing heat bath at negative temperatures are never present.

Our considerations are limited to conditions of thermodynamic equilibrium of systems interacting with an external field and to ideal, quasi-static transformations between equilibrium states. When the formalism is extended to systems that are not strictly thermodynamic (as in vortices in 2D turbulence \cite{onsager1949statistical,pakter2018nonequilibrium}) as well as nonequilibrium or nonextensive conditions other formal negative temperatures are likely to emerge, instead of actual thermodynamic temperatures that cannot be defined \cite{struchtrup2018work}.


\begin{thebibliography}{10}
\bibitem{simplesystems}
	In thermodynamics, simple systems are defined as macroscopic systems that are
	homogeneous, isotropic, and do not interact with electric, magnetic or
	gravitational fields {[24]}.
\end{thebibliography}

\end{document}